\documentclass[aps,pre,twocolumn,floats,amsmath,amssymb,showpacs,floatfix]%
{revtex4}
\usepackage{graphicx}
\usepackage{dcolumn}
\renewcommand{\bbox}[1]{\mbox{\boldmath$#1$}}

\begin{document}
\title{Multiple regimes of diffusion}
\author{B. Mehlig$^{(1)}$, M. Wilkinson$^{(2)}$, V. Bezuglyy$^{(2)}$,
K. Gustavsson$^{(1)}$, K. Nakamura$^{(3,4)}$}
\affiliation{$^{(1)}$Department of Physics, Gothenburg University, 41296
Gothenburg, Sweden\\
$^{(2)}$Department of Mathematics and Statistics,
The Open University, Walton Hall, Milton Keynes, MK7 6AA, England\\
$^{(3)}$Department of Applied Physics, Osaka City University,
Osaka 558-8585, Japan\\
$^{(4)}$Heat Physics Department, Uzbek Academy of Sciences,
28 Katartal Street,
100135 Tashkent,
Uzbekistan}

\begin{abstract}
We consider the diffusion of independent particles experiencing
random accelerations by a space- and time-dependent force as well as
viscous damping. This model can exhibit several asymptotic behaviours, depending
upon the limiting cases which are considered, some of which have been discussed
in earlier work. Here we explore the full space of dimensionless parameters, and introduce
an \lq asymptotic phase diagram' which delineates the limiting regimes.
\end{abstract}

\maketitle

\section{Introduction}
\label{sec: 1}
\begin{figure}[t]
\includegraphics[width=6cm,clip]{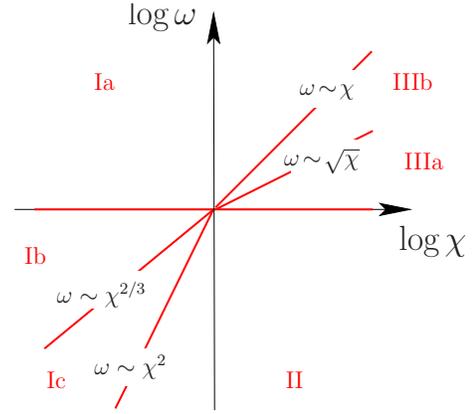}
\caption{\label{fig: 1}
Phase diagram (schematic) for the model (\ref{eq: 1.2}-\ref{eq: 1.4}) summarising
the different dynamical behaviours of Eq.~(\ref{eq: 1.2}) described in sections \ref{sec: 3}-\ref{sec: 5}: I Ornstein-Uhlenbeck, II
generalised Ornstein-Uhlenbeck, IIIb
overdamped minimum tracking, IIIa underdamped minimum tracking.
The Ornstein-Uhlenbeck regime is divided into three regions:
overdamped advection (Ia), underdamped advection (Ib), and
underdamped inertial dynamics (Ic).
}
\end{figure}
\begin{figure}[t]
\includegraphics[width=6cm,clip]{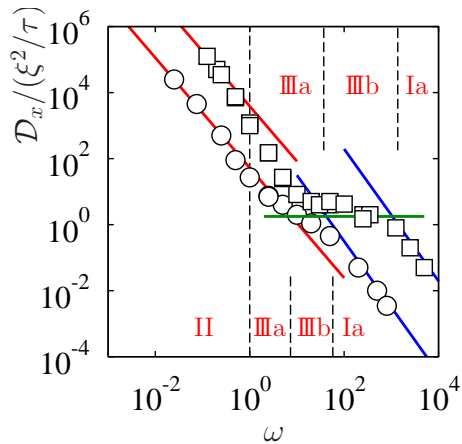}
\caption{\label{fig: 2}
Illustration of multiple regimes of diffusion.
 Evaluation of the diffusion constant ${\cal D}_x$ for constant
 $\chi$ as a function of $\omega$:
 for  $\chi=50$ (circles)
 and $\chi=1250$ (squares). Also shown is the expected behaviour
 in the advective limit, Ia, (blue) and  the expected behaviour in regime II,
 ${\cal D}_x \sim \xi^2/\tau\,\,\chi^{4/3} \omega^{-5/3}$ (red).
Finally, the estimate (\ref{eq: 5.6}) for
${\cal D}_x$ in the minimum-tracking regime is shown (green line).
}
\end{figure}

The position $x$ of a particle subjected to a force which fluctuates
randomly in time $t$ might be expected to undergo diffusion,
in the sense that
\begin{equation}
\label{eq: 1.1}
\lim_{t\to \infty}
\frac{\langle x^2(t)\rangle}{2t}={\cal D}_x
\end{equation}
for some diffusion coefficient ${\cal D}_x$, provided that there
is some damping mechanism preventing the particle from
being accelerated to arbitrarily high velocities. In this paper we determine
the diffusion constant for the
simplest model for this process in one spatial dimension, in which the
equation of motion of the particle is
\begin{equation}
\label{eq: 1.2}
m\dot x=p\ ,\quad
\dot p=-\gamma p+f(x,t)\,.
\end{equation}
Here $x$ and $p$ are particle position and momentum, respectively,
$m$ is the mass, and $\gamma$ is the rate at which
the particle momentum is damped due to viscous drag.
Time derivatives are denoted by dots.
Further, the random forcing $f(x,t)$  is modeled by
a Gaussian random function with zero mean and with correlation function $C(x,t)$,
characterised by a correlation length $\xi$,  correlation
time $\tau$, and of typical size $\sigma$: 
\begin{equation}
\label{eq: 1.3}
\langle f(x,t)\rangle = 0\,,\quad
\langle f(x,t)\,f(x',t')\rangle = {C}(x\!-\!x',t\!-\!t')
\end{equation}
(angular brackets denote averages).
The dynamics of the model (\ref{eq: 1.2}), (\ref{eq: 1.3}) 
is determined by five dimensional 
parameters: $\sigma$, $\tau$, $\xi$, the mass $m$, and $\gamma$. Out of 
these one can
form two independent dimensionless parameters: a dimensionless force
$\chi = \sigma\tau^2 /(m\xi)$ and a dimensionless damping $\omega =
\gamma\tau$. In the following  we explore the full space
of dimensionless parameters $\chi$ and $\omega$.

There is no exact expression for the diffusion
constant for this simple model. However, asymptotic expressions with different
regions of validity are known, depending on which ratios of dimensionless parameters, $\omega$
and $\chi$, approach zero. We show that there are surprisingly many different asymptotic
regimes, which are summarised in an \lq asymptotic phase diagram', Fig.~1.
The axes in this diagram are logarithms of two independent dimensionless parameters of the
model, and a ray from the origin with slope $\nu$ represents a limiting process where
$\chi$ approaches $0$ or $\infty$ with $\omega\sim \chi^\nu$. The phase lines do not indicate
sharp transitions, but rather the boundaries between the domains of 
validity of six different
asymptotic regimes as the limit is taken. Some of the regimes are well understood, but
others are either new or have only been studied recently by the authors of this paper.
It is remarkable that the phase diagram of such a fundamental model for diffusion
processes has not been completely characterised before now.
In all cases the long-time dynamics is diffusive, but in some of the regimes
the stationary distribution of momentum may be strongly non-Gaussian, and the short-time behaviour
may exhibit anomalous diffusion. The diffusion constant ${\cal D}_x$ depends
in different ways on the microscopic parameters in different regimes, as
illustrated in Fig.~\ref{fig: 2}. 

The three-dimensional version of the model defined by Eqs.~(\ref{eq: 1.2})
and (\ref{eq: 1.3}) arises naturally in the study of small particles suspended in a
randomly moving fluid, for which
motion relative to the fluid is determined by viscous drag. In that context the random force is replaced by a
random vector field, which would usually be chosen to be solenoidal, to represent
an incompressible flow. This three-dimensional system has been extensively studied
in certain limits. A significant early contribution is due to Maxey \cite{Max87},  
who analysed the clustering of particles suspended in a turbulent fluid (referred to as \lq preferential concentration').  Ref.~\cite{Wil07} provides an overview of the literature on this problem, and describe recent progress. 

This present paper is the
first to explore the full range of regimes which are possible in limiting cases
of the model, some of which are not realised in fluid-dynamical applications.
We remark that different choices of dimensionless parameters are used in some
other papers: much of the fluid dynamics literature uses the Stokes number
${\rm St}=1/\omega$ as a measure of the damping, and the Kubo number ${\rm Ku}=\chi/\omega$
as a measure of the time scale of fluctuations of the velocity field.

The model also exhibits an interesting effect which involves a phase transition in the
conventional sense. Depending upon the dimensionless parameters of the model, particles with
different initial conditions experiencing the same realisation of the random force 
approach the same trajectory with probability unity. This \lq path coalescence' effect and the
\lq path-coalescence transition' where it disappears, were noted by \citet{Deu85}, who
appears to have been the first to consider this model systematically. In this paper we
also describe the full phase line for the path-coalescence transition, extending
results of \cite{Wil03}.
The critical line for the path-coalescence
transition in the $\chi$-$\omega$ plane is shown in Fig.~\ref{fig: 3}.

The numerical simulations of Eqs.~(\ref{eq: 1.2}), (\ref{eq: 1.3}) described in
this paper were performed with the following choice of correlation function:
\begin{equation}
\label{eq: 1.4}
C(x,t) = \sigma^2 \exp[-x^2/(2\xi^2)-t^2/(2\tau^2)]\,.
\end{equation}
The detailed choice of the correlation function of the force is not
significant, but in regime II, one aspect of the random forcing can make
a qualitative difference. For any choice of the random force, there is
a corresponding potential, satisfying $-\partial V(x,t)/\partial x=f(x,t)$. 
For a generic choice of correlation function, the one-dimensional potential 
$V(x,t)$ corresponding to the force $f(x,t)$
performs a random walk exhibiting increasing fluctuations as $\vert x\vert$ increases.  
We also consider cases where
the particle dynamics is different if the potential $V(x,t)$ is a 
stationary random process. 

The different regimes are illustrated in Fig.~\ref{fig: 4} by 
numerical simulations of the Eqs.~(\ref{eq: 1.2}-\ref{eq: 1.4}).
Shown are the trajectories $x(t)$ of several particles for a given 
realisation $f(x,t)$ of the forcing.

\section{Summary and physical description of the regimes}
\label{sec: 2}

\begin{figure}[t]
\includegraphics[width=6cm,clip]{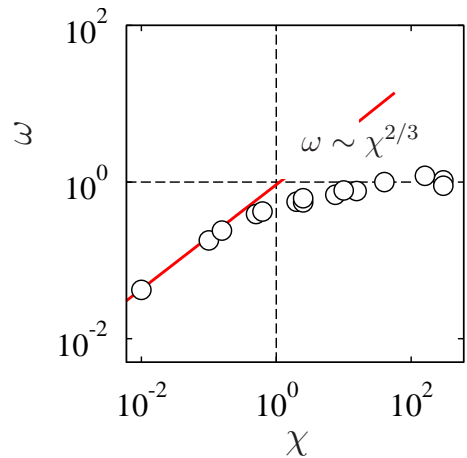}
\caption{\label{fig: 3}
Phase line of the path-coalescence transition.  Results of computer simulations
of (\ref{eq: 1.2}-\ref{eq: 1.4}) for the phase line of the path-coalescence
transition. Also shown is the theoretical result valid for small
values of $\chi$, red line.}
\end{figure}
\begin{figure}
\includegraphics[width=6cm]{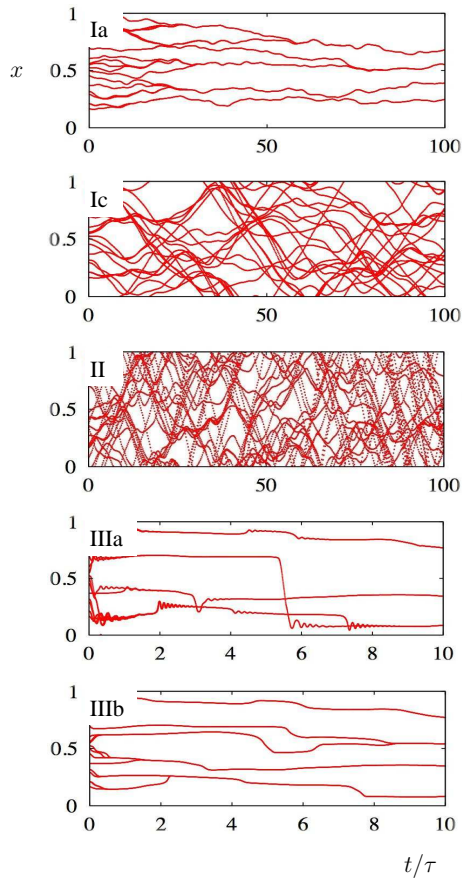}
\caption{\label{fig: 4}
Trajectories $x(t)$ as a function of
$t/\tau$ for $20$ particles
with initial condition $p(0)=0$ and $x(0)$ randomly chosen
in $[0,1]$. The first two pictures (in regimes Ia and Ic) are
 similar to those obtained
 by \citet{Deu85}, and to Fig. 1 in \citep{Wil03}.
The trajectories in region Ib are similar to those in Ia (not shown).
The general difference in the dynamics of regimes IIIa and IIIb is that there are oscillations in regime IIIa, due to smaller damping.
}
\end{figure}

Before describing our results in detail we discuss the physics of the parameter regimes 
of our model. We also mention connections with other work
on the dynamics of randomly forced particles, where some of the regimes of our
model have been studied. 

One limiting case is where the
particles are advected by a random velocity field: there is extensive
literature on this problem and the closely related model of passive scalars \citep{Fal01}.
The advective case corresponds to regime Ia in our model.

Our model (\ref{eq: 1.2}), (\ref{eq: 1.3}) reduces to the well-known
Ornstein-Uhlenbeck process \citep{Orn30} when the position dependence
of the force can be neglected. This is the case when the forcing
is sufficiently weak so that the particle position changes negligibly
within correlation time $\tau$: this condition will be made
more precise below. The Ornstein-Uhlenbeck process is discussed
in standard textbooks (see for example \cite{vKa81}). The regimes Ib,c can both be
analysed by treating the evolution of the momentum as an Ornstein-Uhlenbeck
process. The difference between the two regimes is that regime Ib exhibits 
path coalescence whereas  regime Ic does not.
Despite the fact that regime Ia describes an overdamped process, the formula
for the diffusion constant is the same as for the Ornstein-Uhlenbeck process.
For this reason regimes Ia and Ib,c are treated together in Sec.~\ref{sec: 3}, and 
are referred to as the Ornstein-Uhlenbeck regimes.
 
Our model is also related to stochastic or \lq Fermi' acceleration of
classical particles by random forces, which is used to model the production
of cosmic rays \citep{Fer49}. In these studies the damping 
term (proportional to $\gamma$) is not included in the equation of motion,
and the particle is accelerated to arbitrarily high energies. The 
treatment of the random forcing term in the case where the particle
is rapidly moving, first considered in \cite{Stu66}, is
used in our consideration of regime II. Without damping
the model exhibits anomalous diffusion \citep{Gol91,Ros92}.
For the case where the damping term is included, 
a new dynamical regime was identified in \cite{Arv06,Bez06}
with a non-Maxwellian velocity distributions
(as well as anomalous diffusion at short times, before the damping term starts
to limit the acceleration). This is regime II
in the phase diagram Fig.~\ref{fig: 1}: we call it the \lq generalised
Ornstein-Uhlenbeck regime'. It is discussed in Sec.~\ref{sec: 4}.

In the case where both the damping and the force are strong, the particle
follows a local minimum of the potential $V(x,t)$. This is regime III of the 
phase diagram. This \lq minimum-tracking' regime has not been considered in detail 
in earlier work. It is discussed in Sec.~\ref{sec: 5} below.

In Sec.~\ref{sec: 6} we briefly describe how results for regime II differ for more general 
types of forcing, such as the case where $V(x,t)$ has stationary statistics.

Fig.~\ref{fig: 3} shows numerical results on the path-coalescence transition,
for the choice of correlation function (\ref{eq: 1.4}). For small values of $\chi$,
the phase boundary is in precise agreement with an asymptotic theory discussed
in \cite{Wil03}, which shows that the transition line is determined by the condition
$\omega \chi^{-2/3}\to {\rm const.}$ in the limit as $\chi\to 0$. The data for large
$\chi $ are consistent with the hypothesis 
that the phase line approaches $\omega={\rm const}.$
as $\chi\to \infty$, but we have no compelling argument to support this.

Finally, we comment on the physically accessible range of dimensionless parameters. 
This depends on the nature
of the forcing. In the case of a particle suspended in a turbulent fluid flow with 
velocity field $u(x,t)$, the random forcing is due to viscous drag, and we write
$f(x,t) = m \gamma u(x,t)$. In this case  disturbances in
the fluid velocity field $u(x,t)$ are transported by
$u(x,t)$ itself. This implies that the Kubo number, ${\rm Ku} \equiv u \tau/\xi$
cannot be large (equivalently, $\chi$ cannot be large compared to $\omega$) 
if the random forcing is due to a fluid flow.
In other cases, such as forcing by random electromagnetic fields, the entire phase diagram may be
accessible.

\section{Ornstein-Uhlenbeck regime}
\label{sec: 3}

We term the regimes Ia,b,c in the phase-diagram Fig.~\ref{fig: 1} the \lq Ornstein-Uhlenbeck regimes'. 
In these regimes, 
the particles move so slowly that the distance 
travelled during one correlation time $\tau$ of the random force $f(x,t)$ 
is much smaller than its correlation length $\xi$.
Thus changes in the spatial argument do not contribute significantly to the decorrelation of $f(x,t)$ 
and one may approximate $f(x(t),t)\approx f(x(0),t)$ for times $t$ of the order of or less than the correlation time $\tau$.

Regime I is divided into one overdamped regime, 
regime Ia ($\omega\gg 1$) and two underdamped regimes Ib, c ($\omega \ll 1)$. 

In the overdamped case, the acceleration term in eq. (\ref{eq: 1.2}) is
negligible,  and consequently the particles are advected by the random force, that
is,  $\dot x\approx f(x,t)/(m\gamma)$.
The two regimes Ib and Ic are distinguished by different behaviours of
nearby particles, as can be seen in Fig.~\ref{fig: 4}. 
In regime Ib, initially separate but nearby particle trajectories 
approach each other (path-coalescence regime).  In regime 
Ic, by contrast, initially close particle trajectories do not coalesce.

In the remainder of this section we first briefly describe
the single-particle dynamics (diffusion), and then
summarise what is known about the path-coalescence transition.

To describe spatial diffusion in the overdamped regime,
one integrates the advective equation of motion $\dot x=f(x,t)/(m\gamma)$.
The change in position $\delta x$ during a short time interval $\delta t\gg\tau$ is
\begin{equation}
\label{eq: 3.1}
\delta x=\frac{1}{m\gamma}\int_t^{t+\delta t} {\rm d}t_1\ f\big(x({t_1}),t_1\big)\,.
\end{equation}
In regime I, one may approximate $f(x({t_1}),t_1)\approx f(x({t}),t_1)$, as
pointed out above. In this regime, the fluctuations
of $f$ at a given point $x$ is indistinguishable from the fluctuations of
$f$ along a particle trajectory. In this case it is straightforward
to determine the fluctuations of $\delta x$: 
since the force is assumed to have vanishing mean (\ref{eq: 1.3}), one has 
$\langle\delta x\rangle=0$.
The variance of $\delta x$ is determined by making
use of the fact that $\left<f(t)f(0)\right>$ is small unless $|t|<\tau$.
Evaluating $\langle \delta x^2\rangle$ one finds the standard  
result
\begin{equation}
\label{eq: 3.2}
\langle \delta  x^2\rangle = \frac{2D_0 \delta t}{(m\gamma)^2}\,,
\end{equation}
where $D_0$ is given by
\begin{equation}
D_0=\frac{1}{2}\int_{-\infty}^\infty{\mathrm d}t\left<f(t)f(0)\right>\,.
\label{eq: 3.3}
\end{equation}

The position $x(t)$ at time $t=N\delta t$ of a particle after $N$ microscopic
steps is $x(t)-x(0)=\sum_{i=1}^N\delta x^{(i)}$, where $\delta x^{(i)}$ is the 
increment at the time step number $i$. For the spatial diffusion constant one obtains 
in the usual fashion:
\begin{eqnarray}
\label{eq: 3.4}
{\cal D}_x&=&\lim_{t\to\infty}\frac{\left<(x(t)-x(0))^2\right>}{2t}\\ 
&=&\lim_{t\to\infty}\sum_{i,j=1}^N\frac{\left<\delta x^{(i)}\delta x^{(j)}\right>}{2\delta tN}
=\frac{D_0}{(m\gamma)^2}\,,\nonumber
\end{eqnarray}
where the increments $\delta x^{(i)}$ and $\delta x^{(j)}$ are statistically
independent when $i\neq j$ and $\delta t \gg \tau$.

Consider now the underdamped regimes Ib, c.   
The displacements $\delta p$ of momentum 
(for  a short time interval $\delta t$) obey
\begin{equation}
\label{eq: 3.5}
\delta p=-\gamma\, p\, \delta t+\delta w\,,
\end{equation}
with
\begin{equation}
\label{eq: 3.6}
\delta w=\int_t^{t+\delta t} {\rm d}t_1\ f\big(x({t_1}),t_1\big)\,.
\end{equation}
%
%
In regime I, the force fluctuates sufficiently rapidly
compared to the time scale on which the momentum
relaxes ($\omega \ll 1$) and the change $\delta p$ of momentum
during one correlation time of the force is
small compared to the typical value of $p$. 
Consequently (\ref{eq: 3.5}) is a Langevin equation \citep{vKa81},
describing the standard Ornstein-Uhlenbeck process where
$\delta w$ is Gaussian distributed with
\begin{equation}
\label{eq: 3.7}
\langle \delta w\rangle = 0\,,\quad
\langle \delta  w^2\rangle = 2 D_0 \delta t\,.
\end{equation}
From the corresponding Fokker-Planck equation for
the distribution $P(p,t)$ of momentum $p$
at time $t$ \citep{vKa81}
\begin{equation}
\label{eq: 3.8}
{\partial P\over{\partial t}}={\partial\over{\partial p}}\big(\gamma p
+D_0{\partial\over{\partial p}}\big)P
\end{equation}
one deduces that the steady-state distribution of momentum is Gaussian,
$P(p) \propto \exp (-\gamma p^2/2 D_0)$. 

Also in this underdamped regime the particles diffuse. 
The Fokker-Planck equation (\ref{eq: 3.8}) allows to determine
the correlation function of momentum in
the steady state. The result is \citep{vKa81}:
\begin{equation}
\label{eq: 3.9}
\left<p(t_1)p(t_2)\right>_{\rm steady\,\, state}=
\frac{D_0}{\gamma}\exp(-\gamma|t_2-t_1|)\,.
\end{equation}
This result in turn allows to calculate the spatial diffusion constant:
\begin{eqnarray}
\label{eq: 3.10} 
{\cal D}_x
&=&\lim_{t\to\infty}\frac{\left<x(t)^2\right>}{2t}\\ 
&=&\lim_{t\to\infty}\frac{1}{2tm^2}\int_0^t{\mathrm d}t_1\int_0^t{\mathrm d}t_2 \left<p(t_1)p(t_2)\right>_{\rm steady\,\, state}\nonumber\\ &=&\frac{D_0}{(m\gamma)^2}\,.\nonumber
\end{eqnarray}
By comparing  the results (\ref{eq: 3.4}) and (\ref{eq: 3.10}) one
finds that the spatial diffusion constant is the same
in the over- and the underdamped limits, 
\begin{equation}
\label{eq: 3.11}
{\cal D}_x = \frac{D_0}{(m\gamma)^2} \propto \frac{\xi^2}{\tau}
\frac{\chi^2}{\omega^2}\,.
\end{equation}
Fig.~\ref{fig: 2} shows results of numerical simulations for the diffusion constant of the model (\ref{eq: 1.2}-\ref{eq: 1.4}).
In the Ornstein-Uhlenbeck regime (I) the simulations agree well with Eq.~(\ref{eq: 3.11}).

We now briefly summarise what is known about the path-coalescence
transition which distinguishes regime Ib from Ic.
In the path-coalescing phase (regime Ib), particle trajectories governed by the equation of motion (\ref{eq: 1.2}) coalesce, whereas in regime Ic initially close particle trajectories separate almost surely (see Fig.~\ref{fig: 4}). As was argued in \cite{Wil03},
the maximal Lyapunov exponent $\lambda$ serves as an \lq order parameter' for
the phase transition. The exponent 
describes the rate of change of an infinitesimal separation between two trajectories
\begin{equation}
\label{eq: 3.12}
\lambda=\lim_{t\to \infty} t^{ -1}\log_{\rm e}\Big|\frac{\delta x_t}{\delta x_0}\Big|\,.
\end{equation}
Here $\delta x_0$ is the initial separation of two infinitesimally close trajectories,
and $\delta x_t$ is their separation at time $t$.

In regime Ib, the Lyapunov exponent  is negative while it is positive in regime Ic.
The condition for the phase transition is thus $\lambda =0$.
In regime I, the Lyapunov exponent can be calculated exactly \citep{Wil03}. 
Expressed in terms of the dimensionless parameters $\chi$ and $\omega$,
the phase transition is found to occur at
\begin{equation}
\label{eq: 3.13}
\omega \chi^{-2/3} = {\rm const.}
\end{equation}
Fig.~\ref{fig: 3} shows results of numerical simulations for
the locus of the path-coalescence transition in the $\chi$-$\omega$-plane
for the model (\ref{eq: 1.2}-\ref{eq: 1.4}). In regime I, the transition
line is given by a line of slope $2/3$, as expected from (\ref{eq: 3.13}).

To conclude this section, we briefly discuss the conditions delineating
regime I in Fig.~\ref{fig: 1}. First, in the overdamped
limit, it was assumed that 
the distance covered in time $\tau$ is smaller than $\xi$. 
Estimating the advective velocity by $\sigma/(m\gamma)$ 
we have the condition $\sigma\tau /(m\gamma) \ll \xi$.
Consequently the line distinguishing regimes Ia and IIIb
is given by  the condition that $\omega/\chi$ is of order unity.
Second, in the underdamped limit, $\omega\ll 1$,  the boundary
between regimes Ic and II is parameterised by the condition that
$\omega/\chi^2$ is of order unity. 
This condition is derived in the next section (Eq.~(\ref{eq: 4.17})),
where regime II is discussed.

\section{Generalised Ornstein-Uhlenbeck regime}
\label{sec: 4}

We term the regime II in the phase diagram Fig.~\ref{fig: 1}  the \lq generalised Ornstein-Uhlenbeck regime'.
This regime is defined by 
strong stochastic forcing: the particles move fast and their momenta can be much larger than  $p_0 = m\xi/\tau$. This means that the particle may travel many correlation lengths $\xi$ during one correlation time  $\tau$ and the stochastic force acting on the particle may decorrelate quicker than $\tau$.
Thus the \lq effective correlation time' of the force $\xi m/\sqrt{\langle p^2\rangle}$ can be much smaller than $\tau$.
It is also assumed that the dynamics of regime II is underdamped, i.e. $\omega\ll 1$.

A Fokker-Planck description is adequate in this regime. Define the increment $\delta w$ of the force for a small time interval $\delta t$
\begin{equation}
\delta w=\int_t^{t+\delta t} {\rm d}t_1\ f\big(x({t_1}),t_1\big)\,.
\label{eq: 4.1}
\end{equation}
In regime I, the time dependence of $x(t_1)$ could be neglected, but this approximation
is no longer valid when the forcing is strong.
Instead, one integrates the equation of motion (\ref{eq: 1.2}) to obtain $x(t)=x(0)+\delta x$, where 
\begin{equation}
\label{eq: 4.2}
\delta x\!=\!\frac{1}{m}\int_0^t\!\!{\mathrm d}t' e^{-\gamma t'}\Big[p(0)+\int_0^{t'}\!\!{\mathrm d}t''e^{\gamma t''}f(x(t''),t'')\Big]\,.
\end{equation}
For times smaller than the small time interval $\delta t$, $\delta x$ is small and one may
expand the force around $\delta x=0$. To lowest order in $\delta t$
one obtains
\begin{eqnarray}
\label{eq: 4.3}
\left<\delta w\right>&\approx&\int_0^{\delta t}{\mathrm d}t_1\left<\frac{\partial f}{\partial x}(x(0),t_1)\delta x\right>\nonumber\\
&\approx&\frac{1}{m}\int_0^{\delta t}{\mathrm d}t_1\int_0^{t_1}{\mathrm d}t' \int_0^{t'}{\mathrm d}t''e^{-\gamma(t'-t'')}\\\nonumber&&\hspace*{1cm}\times\left<\frac{\partial f}{\partial x}(x(0),t_1)f(x(t''),t'')\right>
\end{eqnarray}
where we have used that $\left<f(x(0),t)\right>=0$. Since it is assumed that $\omega=\gamma\tau\ll 1$, the exponent $\gamma(t'-t'')\approx 0$ and the major contribution to the integrals from the force correlation is for $|t_1-t''|<\tau$. We get
\begin{equation}
\left<\delta w\right>\approx\frac{\delta t}{2m}\int_{-\infty}^{\infty}{\mathrm d}t\, t\left<\frac{\partial f}{\partial x}(0,0)f(pt/m,t)\right>\,.
\label{eq: 4.4}
\end{equation}
The variance of the displacements (\ref{eq: 4.1})  becomes lowest order in $\delta t$ 
\begin{eqnarray}
\left<\delta w^2\right>&=&\int_0^{\delta t}{\mathrm d}t_1\int_0^{\delta t}{\mathrm d}t_2\left<f\left(pt_1/m,t_1\right)f\left(pt_2/m,t_2\right)\right>\nonumber\\
&\approx&\delta t\int_{-\infty}^\infty{\mathrm d}t \,C(pt/m,t)\,,
\label{eq: 4.5}
\end{eqnarray}
where $C$ is the correlation function of the force (\ref{eq: 1.3}).

Using that the change $\delta p$ of momentum during a short time period is $\delta p=-\gamma p\delta t+\delta w$ together with the first two moments of $\delta w$ 
(Eqs.~(\ref{eq: 4.4}) and (\ref{eq: 4.5})), a
Fokker-Planck equation is obtained using the standard procedure \citep{vKa81}:
\begin{equation}
{\partial P\over{\partial t}}={\partial\over{\partial p}}\left(-v(p)+{\partial\over{\partial p}}D(p)\right)P\,.
\label{eq: 4.6}
\end{equation}
Here the drift- and diffusion-coefficients are
\begin{eqnarray}
v(p)&=&\lim_{\delta t\to 0}\frac{\left<\delta p\right>}{\delta t}=-\gamma p+\frac{\partial}{\partial p}D(p)\nonumber\\
D(p)&=&\lim_{\delta t\to 0}\frac{\left<\delta p^2\right>}{2\delta t}=\frac{1}{2}\int_{-\infty}^\infty{\mathrm d}t\, C(pt/m,t)\,,
\label{eq: 4.7}
\end{eqnarray}
where it was used 
that $\left<\delta w\right>=\delta t{\partial_p}D(p)$.
The above expression for the diffusion constant
was earlier obtained in \cite{Stu66}, and has also been used in \cite{Gol91} and \cite{Ros92}.
Note that when $|p|\ll p_0$,  $D(p)\approx D_0$, which corresponds to the Fokker-Planck equation of the 
standard Ornstein-Uhlenbeck process discussed in Sec.~\ref{sec: 3}.

On the other hand, when $|p|\gg p_0$ we approximate
\begin{equation}
D(p)\!=\!{D_1 p_0\over \vert p\vert}+O(p^{-2}) \,,\,
D_1={m\over 2p_0}\int_{-\infty}^\infty\!\!\!\!\!\! {\rm d}X\ C(X,0) \,.
\label{eq: 4.9}
\end{equation}
For the correlation function (\ref{eq: 1.4}) we obtain
$ D(p)= {D_0}/{(1+p^2/p_0^2)^{1/2}}$, that is $D_1 = D_0$.

If the force is the gradient of a potential $V(x,t)$ with continuous derivatives, $D_1$ vanishes and
 $D(p)\propto  \vert p\vert^{-3}$ provided $V(x,t)$ is sufficiently differentiable. Further, if
the correlation function exhibits a cusp at $t=0$ (an example
is discussed by \cite{Bez06}), we find $D(p)\propto  \vert p\vert^{-2}$.
In general we write
\begin{equation}
\label{eq: 4.10}
D(p) = D_\zeta\,\,\left(\frac{p_0}{|p|}\right)^\zeta
\end{equation}
with $\zeta\geq 0$. In this paper we mainly discuss the case
$\zeta=1$ which is a generic case for a random force [realised by the correlation function (\ref{eq: 1.4})].
But in Sec.~\ref{sec: 6} we briefly mention what is known for other force
models giving rise to $\zeta \neq 1$.
\begin{figure}[t]
\includegraphics[width=6cm]{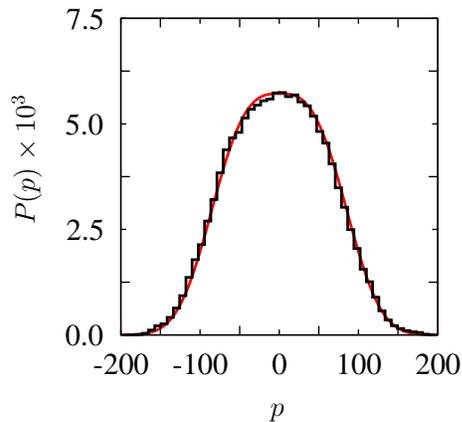}\hfill
\caption{\label{fig: 5} Non-Gaussian distribution of momentum in
phase II (c.f. Fig. \ref{fig: 1}). Shown are results
of numerical simulations of (\ref{eq: 1.2})
for $\chi = 50$ and $\omega =0.01 $
compared with eq. (\ref{eq: 4.12}).  }
\end{figure}
\begin{figure}[t]
\includegraphics[width=7cm]{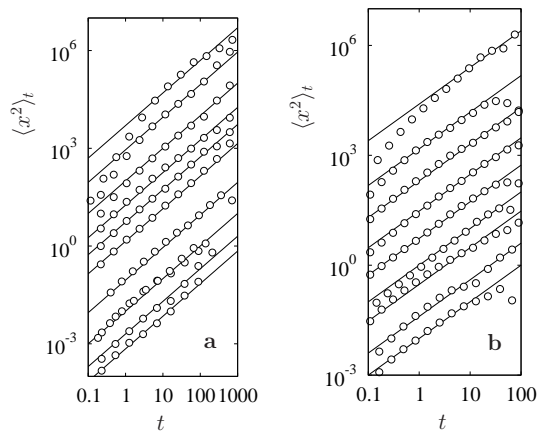}\hfill
\caption{\label{fig: 6} Shows $\langle x^2\rangle_t$ as a function
of $t$; {\bf a} $\chi = 50$  and $\omega = 0.025, 0.075, 0.25, 0.5, 1,
2.5, 50, 200, 500$ and $800$, from top to bottom;
{\bf b} $\chi=1250$ and $\omega = 0.125,
1, 0.5, 2.5, 5, 50, 100, 250, 1250, 2500$ and $5000$, from top to
bottom.
Solid lines
are fits (as judged by the eye) to the diffusion law (\ref{eq: 1.1}). The
corresponding values of ${\cal D}_x$ are shown in
Fig. \ref{fig: 2}.
}
\end{figure}

The steady-state solution $P(p)$ of (\ref{eq: 4.6}) is given by
\begin{equation}
\label{eq: 4.11}
P(p)  = C\,\exp\Big(-\gamma\int_0^p{\rm d}p' \frac{p'}{D(p')}\Big)\,.
\end{equation}
where $C$ is chosen to normalise the distribution.
When $D(p) \approx D_0$, the function $P(p)$ is approximately Gaussian, which corresponds
to the standard Ornstein-Uhlenbeck regime (regimes Ib,c).
When $D(p) = D_1 p_0/|p|$, however,
the distribution $P(p)$ is non-Gaussian (regime II):
\citep{Arv06}
\begin{equation}
\label{eq: 4.12}
P(p)\sim
\exp\Bigg(-\frac{\gamma \vert p\vert^{3}}{3 p_0 D_1}\Bigg)\,.
\end{equation}
This result is compared to results of numerical simulations of 
Eqs.~(\ref{eq: 1.2}-\ref{eq: 1.4}) in Fig.~\ref{fig: 5}.

The generalised Ornstein-Uhlenbeck regime was studied in Ref.~\cite{Arv06}, see also Ref.~\citep{Bez06}.
The Fokker-Planck equation (\ref{eq: 4.6}) was solved as an eigenvalue problem and the propagator to reach momentum $p$ at time $t$ given the initial momentum was found.
From this propagator, the momentum correlation function at equilibrium was calculated
\begin{eqnarray}
\nonumber
\left<p(t')p(t'')\right>_{\mathrm eq.}&=&\frac{\Gamma(4/3)}{3^{1/3}\Gamma(5/3)}\left(\frac{p_0D_1}{\gamma}\right)^{2/3}e^{-2\gamma|t'-t''|}\\&&\times F_{21}\left(\frac{1}{3},\frac{1}{3};\frac{5}{3};e^{-3\gamma |t'-t''|}\right)\,,
\label{eq: 4.13}
\end{eqnarray}
where $F_{21}$ is a hypergeometric function \citep{Abr72}.

From the momentum correlation function (\ref{eq: 4.13}) 
it is possible to calculate the spatial diffusion constant in regime II.
The result is \citep{Arv06}:
\begin{equation}
 \label{eq: 4.14}
 {\cal D}_x\!=\frac{(p_0D_1)^{2/3}}{m^2\gamma^{5/3}}
 \frac{\pi 3^{-5/6}}{2\,\Gamma\!(2/3)^2} F_{32}
 \Big(\frac{1}{3},\frac{1}{3},\frac{2}{3};\frac{5}{3},\frac{5}{3};1\Big)\,,
 \end{equation}
where $F_{32}$ is a hypergeometric function.
Since $D_1 \sim \sigma^2 \tau$, the spatial diffusion constant ${\cal D}_x$
scales as
 \begin{equation}
\label{eq: 4.15}
 {\cal D}_x \sim \frac{\xi^2}{\tau} \chi^{4/3} \omega^{-5/3}
 \end{equation}
as opposed to (\ref{eq: 3.11}).
Fig.~\ref{fig: 2} shows results of numerical simulations
of the diffusion constant for the model (\ref{eq: 1.2}-\ref{eq: 1.4})
in this regime, in good agreement with eq. (\ref{eq: 4.14}).
The diffusion constant ${\cal D}_x$ was numerically determined by
estimating $\langle x^2\rangle_t/(2t)$, according
to Eq.~(\ref{eq: 1.1}). The
corresponding data are shown in Fig.~\ref{fig: 6}.
The dynamics in the generalised Ornstein-Uhlenbeck regime exhibits anomalous diffusion at 
short times \citep{Arv06,Bez06}. Results of numerical experiments exhibiting
anomalous diffusion are given in \cite{Arv06}.

To conclude this section we discuss the conditions under which the results described 
above are applicable. First we discuss the lines in Fig.~\ref{fig: 1} defining the limits of 
regime II. The discussion above assumes that the dynamics in regime II are described by
a Fokker-Planck equation, (\ref{eq: 4.6}), and we must also consider the conditions under
which this equation is applicable. 

Regime II is defined by the condition that the motion is underdamped, $\omega\ll 1$, and
by the condition that the correlation time along the particle trajectory is smaller than 
the correlation time for a static particle. The latter condition defines the transition 
between regimes I and II in the phase diagram Fig.~\ref{fig: 1}. This transition occurs when
$\gamma p_0^2/D_0 $ is of order unity. This fact is most easily
seen by determining the steady-state
distribution $P(p)$ of momentum for the particular
form   (\ref{eq: 1.4}) of the correlation function:
$P(p) = C \exp\big(-\frac{\gamma p_0^2}{3 D_0 }
                \big[(1+p^2/p_0^2)^{3/2}-1\big]\big)$.
In terms of the dimensionless parameters $\omega$ and $\chi$ the
above condition becomes:
\begin{equation}
\label{eq: 4.16}
{\gamma p_0^2\over D_0}\sim {\gamma m^2\xi^2\over{\sigma^2\tau^3}}
=\frac{\omega}{\chi^2}= {\rm const}\,.
\end{equation}
Thus the lines distinguishing the boundaries of regime II in the phase
diagram in Fig.~\ref{fig: 1} are
\begin{equation}
\label{eq: 4.17}
\omega =O(1)\quad \mbox{and}\quad \frac{\omega}{\chi^2} = O(1)\,.
\end{equation}

Consider finally the conditions of validity of (\ref{eq: 4.6}) in regime II.
For a Fokker-Planck description of a stochastic process to be valid,
two necessary conditions must hold
\begin{itemize}
\item {\em Amplitude condition}. The random jumps of the stochastic
variable must be much smaller than its typical size. Therefore
we must require that the change of momentum $\Delta p$ during a correlation time
of the forcing is much smaller than $\sqrt{\langle p^2\rangle }$. This condition 
can be written in the form
\begin{eqnarray}
\frac{\Delta p}{p} \sim \frac{\sigma}{p} \frac{m\xi}{p}
\sim  \frac{\sigma m\xi}{{\langle p^2 \rangle}}
&\sim& \chi \Big(\frac{\omega}{\chi^2}\Big)^{2/3}\ll 1
\,.
\label{eq: 4.19}
\end{eqnarray}
Here we used the fact that 
the correlation time of the force is of the order of
$m\xi/\sqrt{\langle p^2 \rangle}$, and that 
in the steady state 
$\sqrt{\langle p^2\rangle} \sim   
(p_0 D_1/\gamma)^{1/3} \sim p_0 (\chi^2/\omega)^{1/3}$.
The amplitude condition is thus fulfilled provided $\omega^2/\chi \ll 1$.

\item {\em Frequency condition}. The stochastic forcing must fluctuate 
more rapidly than the stochastic variable. 
The correlation time of the force is of the order of
$m\xi/\sqrt{\langle p^2 \rangle}$, and the relaxation time
of $p$ is of the order of $\gamma^{-1}$. We must therefore require that
\begin{equation}
\label{eq: 4.18}
\frac{\gamma \xi m }{\sqrt{\langle p^2 \rangle}}
\sim \omega \Big(\frac{\omega}{\chi^2}\Big)^{1/3}\ll 1\ .
\end{equation}
The frequency condition thus amounts to the same condition
as above.
\end{itemize}

In our model, the amplitude and frequency  conditions 
may not be sufficient to ensure that
(\ref{eq: 4.6}) is valid. The reason is that the fluctuations
experienced by the particle may be influenced by the random force altering 
the trajectory of the particle, so that the trajectory does not explore the random force
field ergodically. 
In addition to the two conditions above the following self-consistency
condition must also be satisfied:
\begin{itemize}
\item {\em Self-consistency condition}. 
It is required that in the steady state the particle moves sufficiently
rapidly so that it is not captured by \lq valleys' in the potential
corresponding to $f(x,t)$:
\begin{equation}
\label{eq: 4.20}
\langle p^2 \rangle/(2m) \gg \xi\sigma\,.
\end{equation}
This condition too corresponds to $\omega^2/\chi\ll 1$.
\end{itemize}

Within the boundaries of region II we have $\omega \ll 1$ 
and $\omega/\chi^2\ll 1$. In this limit the condition $\omega^2/\chi \ll 1$
is always satisfied, so that the Fokker-Planck equation is in fact always applicable
in regime II for the type of random force model with $\zeta=1$. In Sec.~\ref{sec: 6} we shall
see that this may not be true for other values of $\zeta$.

\section{Minimum-tracking regime}
\label{sec: 5}
\begin{figure}[t]
\includegraphics[width=7cm,clip]{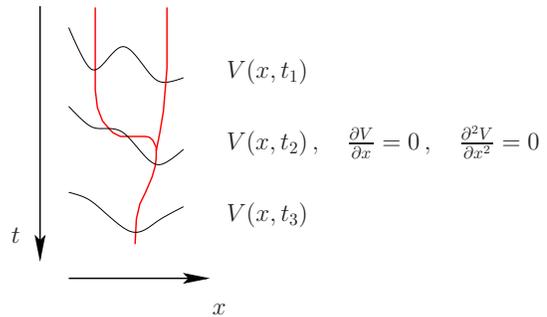}
\caption{\label{fig: 7} Disappearance of a minimum in $V(x,t)$ (black lines)
and corresponding coalescence of particle trajectories (red lines, schematic).}
\end{figure}
We term the regimes III in the phase diagram Fig.~\ref{fig: 1} as the \lq minimum-tracking regimes'.
These regimes are defined by large damping $\omega\gg 1$ and strong stochastic forcing $\chi/\omega\gg 1$.
If the force had been weak, we would be in the advective regime Ia.
When the force becomes large enough,
the particle may get stuck in minima of the potential $V(x,t) =- \int_0^x\!\!{\rm d}x' f(x',t)$ of the force.

The minimum-tracking regimes are divided into two distinct dynamical regimes: under- and overdamped minimum-tracking.
In the underdamped regime IIIa, the particle can oscillate around the potential minimum, whereas in the overdamped regime IIIb such oscillations are quickly damped out (Fig.~\ref{fig: 2}).

In regime IIIb the diffusion constant is estimated as follows. Typically the particles are stuck in minima. As these minima randomly disappear and appear, the particle trajectories jump and may coalesce (Fig.~\ref{fig: 7}).
We argue that the diffusion
constant is given in terms of
the typical size of the valleys ($\xi$) and the rate
at which they disappear.
This rate can be estimated using a
generalisation of the method discussed in \cite{Kac43,Ric45}
for counting the zeroes of a random function $\bbox{F}(\bbox{y})$. 
According to \cite{Kac43}, the density $\varrho$ of zeroes of the random
function $\bbox{F}(\bbox{y})$ is given by
\begin{equation}
\label{eq: 5.1}
\varrho =  \langle \delta(\bbox{F}) |\det {\bf G}|\rangle
\end{equation}
with $G_{\alpha\beta} = \partial F_\alpha/\partial y_\beta$.

Valleys disappear at inflection points of $V(x,t)$, and we thus need to count the joint zeroes of $f$ and $\partial_x f \equiv f'$:
\begin{equation}
\label{eq: 5.2}
\varrho = \langle \delta(f) \delta(f') |\det {\bf G}|\rangle
\end{equation}
with
\begin{equation}
\label{eq: 5.3}
{\bf G} = \left (\begin{array}{ll} f'    & \frac{\partial f}{\partial t}
\\[2mm]
f'' & \frac{\partial f'}{\partial t}
\end{array}\right)\,.
\end{equation}
We need to calculate the average with respect to the set of 
Gaussian random variables $a_i=\{f,f',f'',\partial_t f$,$\partial_t f'\}$. The expectations
of all five random variables vanish, and their covariances $\Sigma_{ij}$
can be expressed in terms of derivatives of the correlation function
$C(x,t)$. For the special case (\ref{eq: 1.4}) we obtain
the covariance matrix
\begin{equation}
\label{eq: 5.4}
\bbox{\Sigma}
= \left(\begin{array}{ccccc}
\sigma^2 & 0 & -\sigma^2/\xi^2 & &  \\
0        & \sigma^2/\xi^2 & 0 & &  \\
-\sigma^2/\xi^2 & 0 & 3\sigma^2/\xi^4 & &\\
   & & & \sigma^2/\tau^2 &\\
   & & & & \sigma^2/(\xi^2\tau^2)
   \end{array}\right)\,.
\end{equation}
Performing the Gaussian average in (\ref{eq: 5.2}), 
we obtain
\begin{equation}
\label{eq: 5.5}
\varrho = \frac{\sqrt{2}}{\pi^2} \frac{1}{\xi\tau}\,.
\end{equation}
Note that this result is independent of $\sigma$.  
For the choice of correlation function (\ref{eq: 1.4}),
the density of minima of $V(x,t)$
for any given time is $(2\pi\xi)^{-1}$. Their
typical separation is thus $2\pi\xi$. Assuming that
particles jump by $2\pi\xi$ when their minima
disappear, we arrive at the following estimate for the diffusion constant
\begin{equation}
\label{eq: 5.6}
{\cal D}_x  = \frac{1}{2} (2\pi)^3  \frac{\sqrt{2}}{\pi^2} \frac{\xi^2}{\tau} = \sqrt{8}\pi \frac{\xi^2}{\tau}
\end{equation}
This expression is in reasonable agreement with the results of numerical simulations shown in Fig.~\ref{fig: 2}.

Now consider the conditions delineating regime IIIb. 
As $\omega$ is increased, regime Ia is entered from IIIb.
The corresponding crossover line was determined
in Sec. \ref{sec: 3}, it is  given by the condition that $\omega/\chi$ is
of order unity.  On the other hand, when
$\omega$ is decreased, regime IIIa is entered, the underdamped
minimum-tracking regime. The corresponding cross-over
line is given by the condition
\begin{equation}
\label{eq: 5.7}
\gamma \omega_{\rm valley} \sim 1\,,
\end{equation}
where $\omega_{\rm valley } = \sqrt{\sigma/(\xi m)}$ 
is the typical frequency of oscillation in the minimum.
The above condition corresponds to
\begin{equation}
\label{eq: 5.8}
\omega^2/\chi = {\rm const.}\,.
\end{equation}

Both minimum-tracking regimes are on the path-coalescence side
of the phase diagram Fig.~\ref{fig: 3}, corresponding to negative
Lyapunov exponent. Calculating this exponent is complicated
by the fact that the minimum-tracking regimes are non-ergodic.
However, in regime IIIa the Lyapunov exponent can be estimated
as follows.

Making use of the fact that the particles are constrained to follow potential 
minima of the random force $f(x,t)$, expand
the potential around a specific minimum at $x_0$ at a given time $t_0$, $V(x,t_0)\approx V(x_0,t_0)+\frac{1}{2}V''(x_0,t_0)(x-x_0)^2$. This gives 
$f=-{\partial_x V}\approx -C(x-x_0)$, where $C=V''(x_0,t_0)>0$.
In the vicinity of $x_0$, the equation of motion (\ref{eq: 1.2}) becomes
\begin{equation}
\label{eq: 5.9}
m\dot x=p,\hspace{2cm}\dot p=-\gamma p-C(x-x_0)\,.
\end{equation}
Linearising for small separations $\delta x$ and $\delta p$ gives
\begin{equation}
\label{eq: 5.10}
\dot {\delta x}=\delta p/m,\hspace{2cm}\dot{\delta p}=-\gamma \delta p-C\delta x\,.
\end{equation}
The dynamics of the separations is determined by the eigenvalues
of the matrix
\begin{equation}
\label{eq: 5.11}
{\bf A}= \left(
\begin{array}{cc}
0   & 1/m \\
-C & -\gamma
\end{array}\right)\,.
\end{equation}
The eigenvalues of ${\bf A}$ are 
\begin{equation}
\label{eq: 5.12}
\lambda_{\pm}=-\frac{\gamma}{2}\pm\frac{\gamma}{2}\sqrt{1-\frac{4C}{m\gamma^2}}\,.
\end{equation}
In regime  IIIa, the discriminant in (\ref{eq: 5.12})
is always negative (this follows from (\ref{eq: 5.8}) and the fact
that $C \sim \sigma/\xi$ which gives $C/m\gamma^2 \sim \chi/\omega^2\gg 1$).
In this regime, the real part of the eigenvalue, which describes the stability of the 
minimum, therefore always takes the value $-\gamma/2$. Because the real part of the eigenvalue 
is constant, we can conclude that the Lyapunov exponent in regime IIIa is $\lambda=-\gamma/2$.

Finally, as $\omega$ is reduced further, eventually region II is entered
when the Lyapunov exponent turns positive. The phase line
in Fig.~\ref{fig: 3} appears to be consistent with the path-coalescence
transition occurring at $\omega = {\rm const.}$. However, we have
not been able to find an argument supporting this observation. Moreover,
we have performed numerical simulations of the path-coalescence
transition at large values of $\chi$ for random forces with correlations
different from (\ref{eq: 1.4}) and have found that the locus of the path-coalescence 
transition sensitively depends on the nature of the fluctuations
of the force. 
This observation is in stark contrast to the boundary between regimes Ib,c where the scaling of the 
zero of the Lyapunov exponent is universal. This latter universality is illustrated
by recent analytical results of \citet{Fal07} (corresponding to regimes Ib,c) for a model
where the force gradient is taken to be a piece-wise
constant function (a so-called \lq telegraph process').

\section{A special case: generalised Ornstein-Uhlenbeck regime}
\label{sec: 6}

Up to now we have considered the simplest case,
where $f(x,t)$ is a Gaussian random function with a generic correlation function.
For such a generic random force its potential function
$V(x,t)$, defined by $f=-\partial V/\partial x$ executes a random walk, exhibiting 
increasing excursions as $\vert x\vert\to \infty$. This generic case corresponds to 
the case where $\zeta = 1$ in (\ref{eq: 4.10}).
For general values of $\zeta$, the long-time diffusion coefficient
and the law of anomalous short-time diffusion in regime II were
calculated by \cite{Bez06}, where it was shown that a force derived from
a statistically stationary potential corresponds to $\zeta=3$.

In this section we briefly discuss  the conditions of validity for 
the Fokker-Planck equation (\ref{eq: 4.6}) for general values of $\zeta$.
The discussion follows that at the end of Sec.~\ref{sec: 4}.

The three conditions to be satisfied are now:

\begin{itemize}
\item {\em Amplitude condition}.
From (\ref{eq: 4.11}) we find that the steady-state
variance of momentum is $\langle p^2 \rangle 
\sim (p_0 D_\zeta/\gamma)^{-2/(2\zeta+1)}$. 
The amplitude condition is therefore:
\begin{eqnarray}
\label{eq:cc9}
\frac{\Delta p}{p} \sim \frac{\sigma}{p} \frac{m\xi}{p}
\sim  \frac{\sigma m\xi}{{\langle p^2 \rangle}}
&\sim& \chi \Big(\frac{\omega}{\chi^2}\Big)^{2/(2\zeta+1)}
\ll 1
\end{eqnarray}
\item{\em Frequency condition}.
In the steady state we require
\begin{equation}
\label{eq: XX}
\gamma \xi m /\sqrt{\langle p^2 \rangle}
\sim \omega \Big(\frac{\omega}{\chi^2}\Big)^{1/(2\zeta+1)} \ll1
\end{equation}
\item{\em Self-consistency condition}. 
For general values of $\zeta$, the self-consistency condition
(\ref{eq: 4.20}) yields
\begin{equation}
\label{eq:cc2}
\frac{2m\xi\sigma}{\langle p^2 \rangle}\sim  \chi 
\Big(\frac{\omega}{\chi^2}\Big)^{2/(2\zeta+1)}\ll 1\,.
\end{equation}
This is the same as (\ref{eq:cc9}).
\end{itemize}

For $\zeta=1$ these three conditions are equivalent and correspond
to the condition $\omega^2/\chi \ll1 $ derived in Sec. \ref{sec: 4}.
But for other values of $\zeta$, the conditions are no longer
equivalent. Condition (\ref{eq:cc2}) can be written as
\begin{equation}
\omega  \ll \chi^{3/2-\zeta}\,.
\end{equation}
This corresponds to a line with slope $-(\zeta-\frac{3}{2})^{-1}$ in Fig.~\ref{fig: 1}.
For $\zeta < 3/2$, this condition is satisfied
everywhere in region II, and therefore does not pose an additional
constraint on the validity of the Fokker-Planck equation
(\ref{eq: 4.6}).
For $\zeta > 3/2$, by contrast, is not satisfied everywhere
in regime II;
a new non-ergodic region appears
in the phase diagram above the dividing line $\omega \chi^{-3/2+\zeta}=$const.
We remark that the condition (\ref{eq: XX}) is always satisfied in regime II.

{\em Acknowledgments.} 
This work was supported by Vetenskapsr\aa{}det (BM), and by
the platform \lq Nanoparticles in an interactive environment' at G\"oteborg university 
(KG, BM).


\begin{thebibliography}{}
\bibitem[Maxey (1987)]{Max87}
{M. R. Maxey}, {\em J. Fluid Mech.}, {\bf 174}, 441-465, (1987).
%
\bibitem[Wilkinson {\em et al.} (2007)]{Wil07}
M. Wilkinson, B. Mehlig, S. \"Ostlund, and K. P. Duncan,
{\em Phys. Fluids}, {\bf 19}, 113303, (2007).
%
\bibitem[Deutsch (1985)]{Deu85} 
J. Deutsch, {\it J. Phys. A: Math. Gen.}, {\bf 18}, 1449, (1985).
%
\bibitem[Wilkinson \& Mehlig (2003)]{Wil03} 
M. Wilkinson and B. Mehlig,
{\it Phys. Rev. E} {\bf 68}, 040101(R), (2003).
%
\bibitem[Falkovich {\em et al.} (2001)]{Fal01} 
G. Falkovich, K. Gawedzki, and M. Vergassola,
{\it Rev. Mod. Phys.}, {\bf 73}, 913, (2001).
%
\bibitem[Ornstein and Uhlenbeck (1930)]{Orn30}
G. E. Ornstein and L. S. Uhlenbeck, {\em Phys. Rev.}, {\bf 36}, 823,  (1930).
%
\bibitem[van Kampen (1981)]{vKa81} N. G. van Kampen,
{\em Stochastic processes in physics and chemistry}, 2nd ed.,
North-Holland, Amsterdam 1981.
%
\bibitem[Fermi (1949)]{Fer49} 
E. Fermi, {\it Phys. Rev.}, {\bf 75}, 1169, (1949).
%
\bibitem[Sturrock (1966)]{Stu66}
P. A. Sturrock, {\em Phys. Rev.}, {\bf 141}, 186, (1966).
%
\bibitem[Golubovic (1991)]{Gol91}
L. Golubovic, S. Feng, and F.-A. Zeng,
{\it Phys. Rev. Lett.}, {\bf 67}, 2115, (1991).
%
\bibitem[Rosenbluth (1992)]{Ros92} M. N. Rosenbluth, {\it Phys. Rev. Lett.}, {\bf 69}, 1831, (1992).
%
\bibitem[Arvedson {\em et al.} (2006)]{Arv06}
E. Arvedson, B. Mehlig, M. Wilkinson, and K. Nakamura, {\em Phys. Rev. Lett.}, {\bf 96}, 030601, (2006).
%
\bibitem[Bezuglyy {\em et al.} (2006)] {Bez06} 
V. Bezuglyy, B. Mehlig, M. Wilkinson, K. Nakamura, and
E. Arvedson, {\it J. Math. Phys.}, {\bf 47}, 073301, (2006).

\bibitem[Abramowitz (1972)]{Abr72}M. Abramowitz and I. A. Stegun, {\it Handbook of Mathematical Functions}, 9th edition, Dover, New York, (1972).

\bibitem[Kac (1943)]{Kac43}
M. Kac, {\it Bull. Am. Math. Soc.}, {\bf 49}, 314, (1943).

\bibitem[Rice (1945)]{Ric45} 
S.O. Rice, {\em Bell System Tech. J.}, {\bf  25}, 46-156, (1945).

\bibitem[Falkovich {\em et al.} (2007)]{Fal07}
G. Falkovich, S. Musacchio, L. Piterbarg, and M. Vucelja,
{\em Phys. Rev. E}, {\bf 76}, 026313, (2007). 


%
%
\end{thebibliography}
\end{document}